# On the nature of the outward pressure in the Universe

**Antonio Alfonso-Faus**

E.U.I.T. Aeronáutica, Plaza Cardenal Cisneros s/n, 28040 Madrid, SPAIN

E- mail: aalfonsofaus@yahoo.es

**Abstract.** Plasma Physics defines the concept of the Debye length. This gives the size of a volume such that from outside it the inside electrical charges, positive and negative, screen each other. Given the enormous electrical potential that would develop with no screening, we conjecture that the size of the Universe, as given by the speed of light c and its age t, ct, cannot be very much lower than the Debye length. We present a cosmological principle of complete interaction: any interaction having a cosmological range is self-contained. The strength of this interaction has a value just to interact at a scale of order ct. Then the Debye length of the Universe must be of the order of its size ct. It turns out that it is about 6 times larger. Hence inside the volume of size ct there is an outward electrical pressure due to the lack of complete screening of the electrical charges. We suggest that these charges are related to the vacuum fluctuations. To the question of why the Universe does not collapse under its own gravitational attraction we analyze the possibility to answer this by the effect of the repulsive forces of these electrical charges. Evidence in support of the idea that space is not expanding is growing more and more in the scientific literature. As a frame of work we consider the model of a constant size Universe that implies a constant equilibrium between the gravitational ($\sim GM^2/R^2$) and the electrical forces ($\sim Q^2/R^2$). We find a large number, $5 \times 10^{60}$ that converts Planck units of mass, length, time, charge $e$ and angular momentum $\hbar$, into the mass M, size ct, age t, total charge Q and angular momentum of the Universe as in a scale-like way. It defines a black hole of mass M $\approx 10^{56}$gr., size $10^{28}$ cm., characteristic time $5 \times 10^{17}$ sec, charge Q $\approx 3.3 \times 10^{60}$ $e$ and maximum angular momentum Ħ $\approx 10^{120}$ $\hbar$ which is our Universe. We are inside it, a part of it. The solution to the Einstein cosmological equations, including an electrical pressure due to the charge Q, is in agreement with the value of the cosmological parameters currently reported.

## 1. - Introduction: plasma physics.

Plasma physics deals with an ionized gas with positive ions and negative electrons forming a cloud that may also include neutral atoms. The degree of ionization refers to the percentage of free charges. It depends on the temperature of the ionized gas: the higher the temperature the higher the degree of ionization. A characteristic length in the plasma is the Debye length $\lambda_D$ given by [1]:

$$\lambda_D^2 = \frac{kT}{4\pi n_0 e^2} \qquad (1)$$

Here kT is the average kinetic energy of the electrons of charge *e* and number density $n_0$. The Debye length in (1) defines the minimum volume of gas such that inside it the net charge is nearly zero due to screening. Of course there are oscillations at the plasma frequency $\omega_p$ given by:

$$\omega_p^2 = \frac{4\pi n_0 e^2}{m} \qquad (2)$$

Let us take a reference model for the Universe such as the spherical one, closed, finite and unbounded with size ct, the speed of light times the age of the Universe. This product must be a constant [2] and therefore a Universe of constant finite size in equilibrium at any instant must continue in equilibrium if no important perturbation appears. In fact the stability of the Einstein static Universe has been analysed [3] and found to be the case for certain conditions.

Any interaction of cosmological importance must have a range R of the order of ct. If R were >> ct the interaction would then be too strong to be confined inside the Universe. If R were << ct the interaction would be too weak to have a cosmological significance, contrary to the initial assumption. This heuristic argument forces the value of R to be of the order of ct. We may call this effect a cosmological principle of complete interaction. We will prove that the Debye length of the Universe is of the order of ct.

Most of the content of the Universe must be in a plasma state. The stars, the solar and stellar winds (of density a few particles per cc. at the Earth's orbit) which are mostly ionized hydrogen gas, protons and electrons. Of course the dark matter and dark energy particle constituents of the Universe are not yet known. Following the argument in the previous paragraph the Debye length for the Universe $\lambda_D$ must be of order ct. Then there must be a characteristic charge Q in the Universe, implying a number of electrons $N_e$ = Q/*e*. We conjecture that it is related to vacuum fluctuations because this number is the same as the maximum number of Planck's fluctuations given by the ratio of the mass of the Universe M to the Planck's unit of mass.

## 2. Cosmological electric charge Q. The fine structure constant revisited

We will now show that the Debye length of the Universe $\lambda_D$ is near 6 times larger than its size ct. The immediate conclusion is that there is a net charge $Q = N_e\,e$ inside the Universe.

We need the term kT in (1) expressed as an average energy. The charge Q we are dealing with is supposed to be related to the vacuum fluctuations and it is inside the whole Universe. Using Planck's mass $m_*$ we must have

$$kT \approx m_* c^2 = \left(\frac{\hbar c^5}{G}\right)^{1/2} \tag{3}$$

On the other hand the fine structure constant $\alpha$ is given by

$$\alpha = \frac{e^2}{4\pi\hbar c} = \frac{1}{137} \tag{4}$$

and can be expressed in the following way: using $\hbar \approx m_p c r_p$ where $m_p$ is the characteristic mass of a fundamental particle of size $r_p$ and contributing significantly to the mass of the Universe, the number of these particles is $N_p \approx 10^{80}$ and its size $r_p \approx 10^{-12}$ cm. Also the size of the Universe we use is ct $\approx 10^{28}$ cm so that ct/$r_p \approx 10^{40}$. Then

$$10^{120}\hbar \approx (10^{80} m_p)c(10^{40} r_p) \approx Mc(ct) \tag{5}$$

We may define maximum angular momentum content for the Universe $\mathrm{H} \approx 10^{120}\,\hbar$. As presented elsewhere [4] the total number of gravity quanta in the Universe is about $10^{120}$ which means that each gravity quanta may have a spin of order ℏ.

From (4) and (5) we get

$$\frac{137}{4\pi}10^{120}e^2 = 10^{120}\hbar c \approx Mc^2(ct) \tag{6}$$

where M is the mass of the Universe. Since the charge Q is spread throughout the whole universe we have

$$\frac{Q^2}{ct} \approx Mc^2 \tag{7}$$

and using (6) we get

$$Q \approx \left(\frac{137}{4\pi}\right)^{1/2} 10^{60} e \approx 3.3 \times 10^{60} e \tag{8}$$

Then the number of charges $e$ due to fluctuations inside the Universe must be

$$N_e \approx 3.3 \times 10^{60} \tag{9}$$

This number has a deep meaning. It converts a Planck's fluctuation into the Universal physical scales. The number density of charges is then $n_0 = N_e / 2\pi^2 (ct)^3 \approx 1.7 \times 10^{-25}$ particles per cc. or $6 \times 10^{24}$ cubic centimetres per particle, i.e. one particle per volume of size 1000 Km.

Now for the Debye length (1) we get using (3), (8) and (9)

$$\lambda_D^2 = \left(\frac{\hbar c^5}{G}\right)^{1/2} \frac{2\pi^2 (ct)^3}{4\pi N_e} \frac{137}{4\pi \hbar c} =$$

$$= \frac{137}{8} (ct)^2 \frac{1}{N_e} \frac{ct}{(G\hbar/c^3)^{1/2}} \approx 32 (ct)^2 \tag{10}$$

The final result is

$$\frac{\lambda_D}{ct} \approx 5.7 \tag{11}$$

hence inside the Universe of size ct there must be a net charge Q that is responsible for an outward pressure that counteracts the inward gravitational attraction. The Universe can be considered as a black hole with mass M ≈ $10^{56}$gr., size $10^{28}$cm. charge Q ≈ $3.3 \times 10^{60}$ *e* and maximum angular momentum Ħ ≈ $10^{120}$ ℏ. In fact the equivalent Compton wavelength for the Universe with an equivalent Planck´s constant Ħ gives Ħ/Mc ≈ ct which is the similar condition used to define the Planck´s quantum black hole using G, ℏ and c. It is a question of scale as presented elsewhere [5].

## 3. The large number $N_e$

In 1937 Dirac [6] proposed his large number hypothesis (LNH) suggested by the numerical coincidences of two large numbers: the ratio of electric and gravitational forces between an electron and a proton, which is of the order of $10^{40}$, and the ratio of the age of the Universe to the time light takes to travel the size of a fundamental particle. Dirac generalized this result to say that any number which is a power of $10^{40}$ will be time dependent to the same power, which constitutes his LNH. He kept all quantum properties constant and left G to vary as 1/t. For the number of particles in the Universe, $N_p$ ≈ $10^{80}$, this hypothesis implies $N_p$ ≈ $t^2$, i.e. particle creation with time. The results of many experimental observations do not support this hypothesis, Uzan [7]. And no creation of particles is observed either. The conclusion is that $N_p$ appears to be constant and therefore the Dirac LNH is not confirmed by experiments. He expressed the idea that very large dimensionless universal constants cannot be pure mathematical numbers, and therefore that they must not occur in the basic laws of physics.

He considered that these numbers were large because they are time varying, and the Universe is old. But this idea may be pursued in a brand new way. We have seen elsewhere [2] that at a cosmological scale all dimensionless universal constants are of order one, which basically is Dirac´s first idea. Therefore at a universal scale there are no large numbers at all, as Dirac thought. This means that the dimensionless cosmological numbers

$$\Omega_m = \frac{8\pi}{3}\frac{G\rho}{H^2} \;,\quad \Omega_e = \frac{8\pi}{3}\frac{Gp_e}{c^2 H^2} \;,\quad \Omega_k = \frac{kc^2}{R^2 H^2} \;,\quad \Omega_\Lambda = \frac{\Lambda c^2}{3H^2}$$

(12)

used in the Einstein cosmological equations, are all constants of order one (here H is the Hubble parameter). So is the cosmological constant in the cosmological units we choose elsewhere [2]. We conclude that the large numbers (or very small ones) come in when considering smaller scales, much smaller than the whole Universe. In particular they come in at the quantum scale, when considering the quantum black holes defined by the three parameters G, ℏ and c known as the Planck´s units. And we include here the charge of the electron e. The Planck's units are

$$m_* = \left(\frac{\hbar c}{G}\right)^{1/2} \approx 2.2 \times 10^{-5} \, gr$$

$$l_* = \left(\frac{G\hbar}{c^3}\right)^{1/2} \approx 1.6 \times 10^{-33} \, cm \quad (13)$$

$$t_* = \left(\frac{G\hbar}{c^5}\right)^{1/2} \approx 5.4 \times 10^{-44} \, \sec$$

and can be multiplied by the large number in (9) to get

$$N_e m_* \approx M \approx 10^{56} \, gr$$

$$N_e l_* \approx ct \approx 10^{28} \, cm$$

$$N_e t_* \approx t \approx 4 \times 10^{17} \, \sec \quad (14)$$

$$N_e e = Q$$

$$N_e^2 \hbar \approx \hbar\!\!\!\!\!-$$

We conclude here that the quantum black hole given by the Planck's units is converted in a scale-like way by the large number $N_e$ into the whole Universe as a cosmological black hole with parameters M, Q and $\hbar\!\!\!\!\!-$.

**4. Cosmological equations with positive electric pressure**

The Einstein cosmological equations are

$$\left(\frac{\dot a}{a}\right)^2 + \frac{2\ddot a}{a} + 8\pi G \frac{p}{c^2} + \frac{kc^2}{a^2} = \Lambda c^2$$

(15)

$$\left(\frac{\dot a}{a}\right)^2 - \frac{8\pi}{3} G\rho + \frac{kc^2}{a^2} = \frac{\Lambda c^2}{3}$$

The pressure term *p* now contains two parts: one is the classical equation of state $p_m = w\rho c^2$ and the other the electrical pressure $p_e \approx Q^2/a^4$. We choose the case of a static Universe [2] and [3]. Using the dimensionless parameters Ω from (12) the two equations in (15) transform to:

$$3w\Omega_m + \Omega_e + \Omega_k = 3\Omega_\Lambda$$
$$-\Omega_m + \Omega_k = \Omega_\Lambda \tag{16}$$

The solution to the Einstein cosmological equations (16) is in line with:

$$w = -1$$
$$\Omega_k = 1$$
$$\Omega_m = 1/3$$
$$\Omega_\Lambda = 2/3 \tag{17}$$
$$\Omega_e = 2$$

The solution found $\Omega_e = 2$ is in agreement with our assumption in (7), that the charge Q has a universal range.

**5. Conclusion**

The Debye length of the Universe turns out to be somewhat larger than its size ct. Therefore there is a net charge Q in the Universe that we have calculated as $5 \times 10^{60} e$. This implies an outward electrical pressure (an electric potential density) that counteracts the gravitational attraction. The Universe does not collapse under its own gravitational attraction. It may be due to the repulsive force of these electrical charges.

A large number, $5 \times 10^{60}$, converts Planck's units of mass, length, time, the charge e and the spin ℏ into the mass M, size ct, age t, total charge Q and maximum angular momentum Ħ for the Universe. It defines a scaled black hole of mass $M \approx 10^{56}$ gr., size $10^{28}$ charge $Q \approx 3.3 \times 10^{60}$ e and angular momentum $Ħ \approx 10^{120}$ ℏ. The solution to the Einstein cosmological equations, when including the charge Q, is in agreement with the numerical values of the cosmological parameters currently observed.